**Title:** Duke Spleen Data Set: A Publicly Available Spleen MRI and CT dataset for Training Segmentation

**Authors:** Yuqi Wang[1], Jacob A. Macdonald[2], Katelyn R. Morgan[2], Danielle Hom[2], Sarah Cubberley[2], Kassi Sollace[2], Nicole Casasanto[2], Islam H. Zaki[2], Kyle J. Lafata[1,2,3], Mustafa R. Bashir[2,4,5]

**Author Affiliations:** [1]Department of Electrical and Computer Engineering, Duke University, [2]Department of Radiology, Duke University, [3]Department of Radiation Oncology, Duke University, [4]Department of Medicine, Duke University, [5]Center for Advanced Magnetic Resonance Development, Duke University

Introduction
Abnormal spleen volume has been associated with chronic liver disease and portal hypertension [1-2]. Spleen volumetry on cross-sectional imaging is feasible in individual cases, but too time-consuming for broad application. Automatic spleen segmentation could be more efficient but is a challenging task due to variability in the shape and texture of the spleen. In recent years, deep learning algorithms have been used in biomedical image segmentation tasks and perform well for segmenting different organs and tissues, [3] but training a segmentation deep neural network requires a suitable dataset. Ideally, models should be able to perform segmentation on different image types (modality, scanner vendor, imaging plane, and image contrast) and on both normal and abnormal spleens in order to be as broadly applicable as possible.

It is non-trivial to collect data and perform manual segmentation, and for this application, not many spleen segmentation datasets are publicly available [4]. The spleen segmentation in the medical segmentation decathlon dataset [5] that comprises 61 CT scans is the only single-organ segmentation dataset for spleen to the best of our knowledge. Spleen segmentation data can also be found in two multi-organ segmentation datasets: CHAOS - Combined (CT-MR) Healthy Abdominal Organ Segmentation Challenge Data [6] and Gibson's dataset [7]. CHAOS includes only 20 MR T1-Dual and T2 SPIR sequences with reference segmentations. Gibson et al. merged the Cancer Image Archive (TCIA) Pancreas-CT data set [8-10] and the Beyond the Cranial Vault (BTCV) Abdomen data set [11-12] to form a multi-organ dataset with 90 abdominal CT scans.

Spleen volumetry is most clinically interesting in patients who may have abnormally shaped and sized spleens, commonly patients with chronic liver disease and portal hypertension. These patients often also display background confounders, including abdominal varices and ascites. To our knowledge, none of the publicly available data sets include these confounding features.

To facilitate the development of spleen segmentation models and efficient spleen volumetry, we provide the Duke Spleen Data Set (DSDS) consisting of 29 axial CT post-contrast series, 40 coronal MRI SSFSE series, and 40 axial MRI opposed phase series, in total 109 CT and MRI volumes. The multi-modality, multi-vendor, multi-planar, multi-contrast nature of the data set as well as the underlying disease states altering spleen appearance are intended to support the development of models robust to image type and presence of disease.

Materials and Methods
*Patients and Image Acquisition*
This data release has been approved by the Duke Office of Clinical Research at Duke University. The dataset (109 volumes; 6322 images in total) was collected from 69 patients (47 male, 22 female; age 41-88) randomly selected from prior studies in patients with known or suspected chronic liver disease, including patients with and without radiologic evidence of cirrhosis and/or portal hypertension. They include images from two modalities (MRI, CT), two common vendors (GE, Siemens), field strengths (1.5, 3.0 T), and normal and abnormal spleen volumes (representative examples in Figure 1). The DSDS covers a wide range of spleen volumes (104.36 – 2025.07 $cm^3$) including both normal and abnormal spleen volumes. A histogram summary is shown in Figure 2.

The 109 volumes can be classified as three types with regard to modality and image plane: 29 axial CT post-contrast series, 40 coronal MRI SSFSE series, and 40 axial MRI opposed phase series. Selected parameters for each image series type are summarized in Table 1.

*Spleen Segmentation*

Manual segmentation was performed by one of four imaging core lab technologists trained in spleen segmentation. Initial segmentations were reviewed and, when necessary, adjusted by one fellowship-trained abdominal Radiologist with 12 years post-fellowship experience in abdominal MRI to yield final segmentations.

The segmentations are provided as separate binary DICOM images, in which 0 represents the background while 1 represent the spleen. An overview of the three image types is shown in Figure 1.

Resulting Dataset

The Duke Spleen Data Set (DSDS) provides 109 anonymized MRI and CT series (6,322 unique images) that can be used to train spleen segmentation and acquire spleen volumes. This dataset can be downloaded from Zenodo (10.5281/zenodo.7636640).

Discussion

We have provided a multi-modality, multi-vendor, multi-planar, multi-contrast dataset of 109 CT and MRI series with manually segmented spleen references, which may serve as a useful data set for training segmentation models, or as a helpful supplement to the publicly available datasets of segmentation masks for normal spleen. The entire data set can be used to generalize spleen segmentation models, or subsets of the data set may be used depending on the desired application. Additionally, the data set fills a gap of single-organ spleen segmentation datasets, including the first of its kind using MR imaging and including spleens of abnormal shape and size. We hope this dataset can help address the challenges in obtaining clinical imaging data with high-quality manual segmentations.

The DSDS dataset was collected from a single institution, which may have resulted in bias related to the local characteristics of imaging protocols or patient population. It also includes data from the two major international imaging system vendors, but not others, and further only includes a limited number of cases for feasibility reasons. Another potential bias source could be the small number of people who performed manual segmentation and the associated variabilities in spleen circumscription due to workstation preferences (window level, screen resolution, etc.) and user edge detection.

References

1. Prassopoulos P, Daskalogiannaki M, Raissaki M, et al. Determination of normal splenic volume on computed tomography in relation to age, gender and body habitus. Eur Radiol. 1997; 7(2):246–8.
2. Bolognesi M, Sacerdoti D, Bombonato G, et al. Change in portal flow after liver transplantation: effect on hepatic arterial resistance indices and role of spleen size. Hepatology. 2002; 35(3):601–8. [PubMed: 11870373]
3. Isensee F, Jaeger PF, Kohl SA, Petersen J, Maier-Hein KH. nnU-net: a self-configuring method for deep learning-based biomedical image segmentation. Nat Methods. 2020;18:1–9.
4. Rickmann AM, Senapati J, Kovalenko O, Peters A, Bamberg F, Wachinger C. AbdomenNet: deep neural network for abdominal organ segmentation in epidemiologic imaging studies. BMC Med Imaging. 2022 Sep 17;22(1):168. doi: 10.1186/s12880-022-00893-4. PMID: 36115938; PMCID: PMC9482195.
5. Antonelli M, Reinke A, Bakas S, Farahani K, Kopp-Schneider A, Landman BA, Litjens G, Menze B, Ronneberger O, Summers RM, van Ginneken B, Bilello M, Bilic P, Christ PF, Do RKG, Gollub MJ, Heckers SH, Huisman H, Jarnagin WR, McHugo MK, Napel S, Pernicka JSG, Rhode K, Tobon-Gomez C, Vorontsov E, Meakin JA, Ourselin S, Wiesenfarth M, Arbeláez P, Bae B, Chen S, Daza L, Feng J, He B, Isensee F, Ji Y, Jia F, Kim I, Maier-Hein K, Merhof D, Pai A, Park B, Perslev M, Rezaiifar R, Rippel O, Sarasua I, Shen W, Son J,

Wachinger C, Wang L, Wang Y, Xia Y, Xu D, Xu Z, Zheng Y, Simpson AL, Maier-Hein L, Cardoso MJ. The Medical Segmentation Decathlon. Nat Commun. 2022 Jul 15;13(1):4128. doi: 10.1038/s41467-022-30695-9. PMID: 35840566; PMCID: PMC9287542.

6. Kavur AE, Selver MA, Dicle O, Barış M, Gezer NS. (2019). CHAOS - Combined (CT-MR) Healthy Abdominal Organ Segmentation Challenge Data (Version v1.03) [Data set]. Zenodo. http://doi.org/10.5281/zenodo.3362844

7. Gibson E, Giganti F, Hu Y, Bonmati E, Bandula S, Gurusamy K, Davidson B, Pereira SP, Clarkson MJ, Barratt DC. Automatic multi-organ segmentation on abdominal CT with dense v-networks. IEEE Transactions on Medical Imaging, 2018. doi:10.1109/TMI.2018.2806309

8. Roth HR, Farag A, Turkbey EB, Lu L, Liu J, and Summers RM. (2016). Data From Pancreas-CT. The Cancer Imaging Archive. http://doi.org/10.7937/K9/TCIA.2016.tNB1kqBU

9. Roth HR, Lu L, Farag A, Shin H-C, Liu J, Turkbey EB, Summers RM. DeepOrgan: Multi-level Deep Convolutional Networks for Automated Pancreas Segmentation. N. Navab et al. (Eds.): MICCAI 2015, Part I, LNCS 9349, pp. 556–564, 2015. http://arxiv.org/pdf/1506.06448.pdf

10. Clark K, Vendt B, Smith K, Freymann J, Kirby J, Koppel P, Moore S, Phillips S, Maffitt D, Pringle M, Tarbox L, Prior F. The Cancer Imaging Archive (TCIA): Maintaining and Operating a Public Information Repository, Journal of Digital Imaging, Volume 26, Number 6, December 2013, pp 1045-1057. http://doi.org/10.1007/s10278-013-9622-7

11. Xu Z, Lee CP, Heinrich MP, Modat M, Rueckert D, Ourselin S, Abramson RG, and Landman BA, Evaluation of six registration methods for the human abdomen on clinically acquired CT, IEEE Trans. Biomed. Eng., vol. 63, no. 8, pp. 1563–1572, 2016.http://doi.org/10.1109/TBME.2016.2574816

12. Landman BA, Xu Z, Igelsias JE, Styner M, Langerak TR, and Klein A, MICCAI multi-atlas labeling beyond the cranial vault - workshop and challenge, 2015, https://doi.org/10.7303/syn3193805

13. Linguraru MG, Sandberg JK, Jones EC, Summers RM. Assessing splenomegaly: automated volumetric analysis of the spleen. Acad Radiol. 2013;20(6):675-684. doi:10.1016/j.acra.2013.01.011

*Table 1: Sequence parameters for each image series type.*

| Sequence | Number of series | Number of slices | Slice thickness [mm] | Pixel Spacing [mm] | Rows | Columns | Repetition time [ms] | Echo time [ms] | Flip angle [°] | Pixel bandwidth [kHz] | Field strength [T] |
|---|---|---|---|---|---|---|---|---|---|---|---|
| **Axial CT post-contrast** | 29 | 34 - 150 | 2.5 - 5.0 | 0.63-0.98 | 512 | 512 | | | | | |
| **Coronal MRI SSFSE** | 40 | 20 - 45 | 6.0 - 7.0 | 0.78 - 1.72 | 256 - 512 | 256 - 512 | 800 - 2769.23 | 78 - 101.664 | 90 - 180 | 244 - 700 | 1.5 - 3.0 |
| **Axial MRI opposed phase** | 40 | 26 - 80 | 4.0 - 8.0 | 0.63 - 1.72 | 256 - 640 | 256 - 640 | 4.06 - 260 | 1.23 - 2.39 | 9 - 90 | 195 - 1040 | 1.5 - 3.0 |

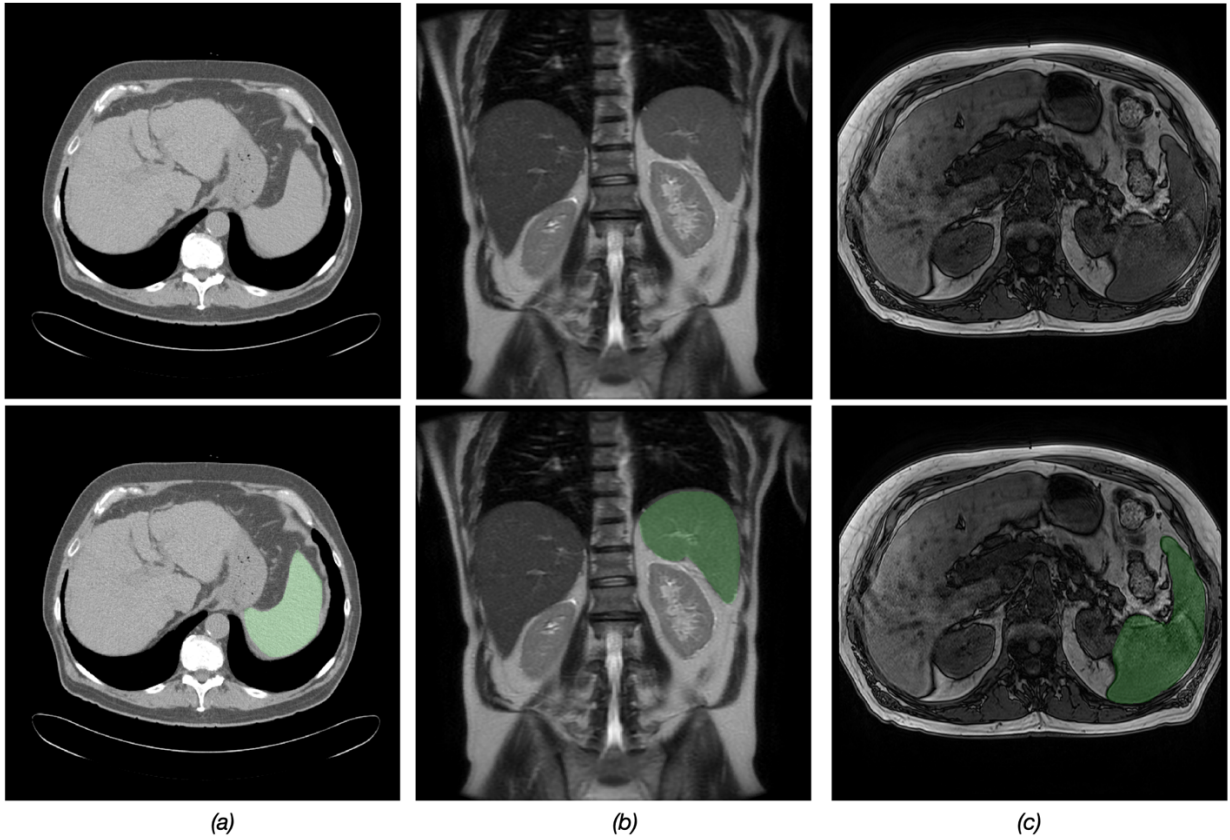

Figure 1: Representative cases of the three series types in the DSDS with the overlaid manual segmentation in green: (a) axial CT post-contrast, (b) coronal MRI SSFSE, and (c) axial MRI opposed phase.

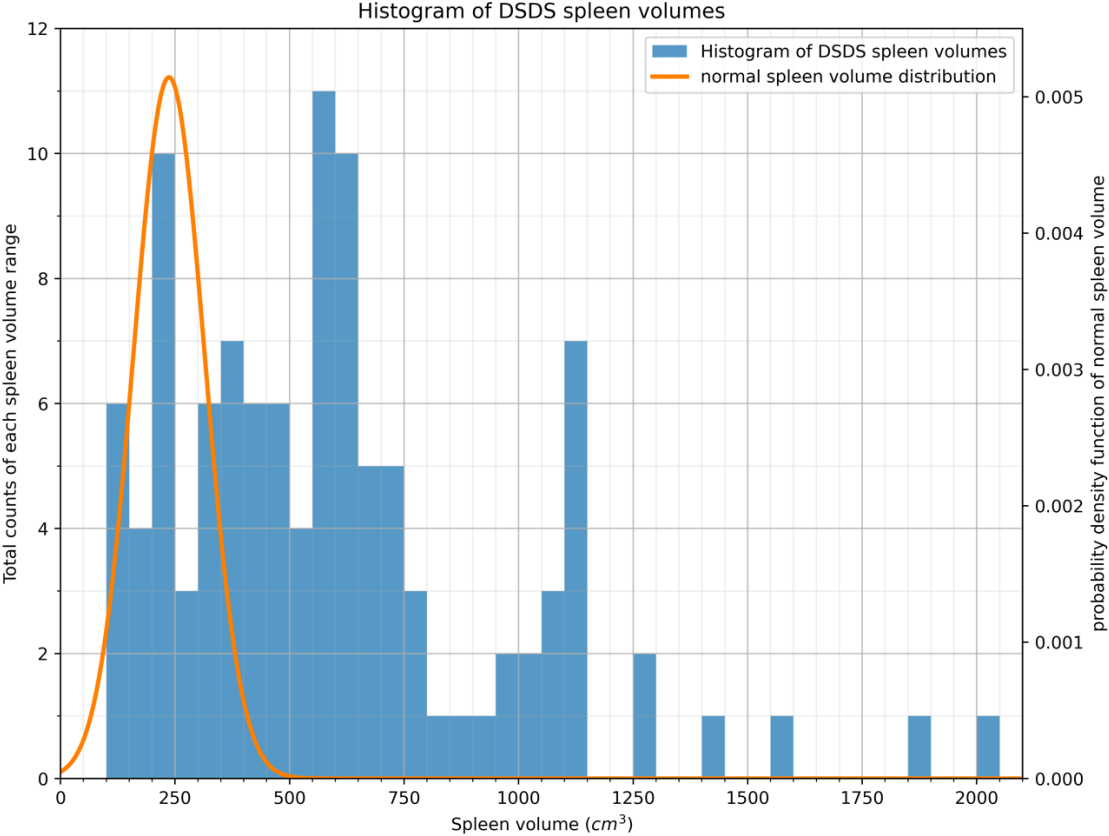

*Figure 2: The histogram of the spleen volumes in DSDS. The blue bars are the histogram of the DSDS spleen volumes, and the overlaid orange curve shows the normal spleen volume range reported by Linguraru et al. [13] measured by 45 CT scans.*